\documentclass[cameraready]{Interspeech}
\title{SLICE: Speech Enhancement via Layer-wise Injection of Conditioning Embeddings}
\author[affiliation={1}, orcid=0009-0002-3994-7018]{Seokhoon}{Moon}
\author[affiliation={2}, orcid=0009-0006-4946-6305]{Kyudan}{Jung}
\author[affiliation={2}, orcid=0000-0003-1071-4835, correspondingauthor]{Jaegul}{Choo}

\address{
    $^1$ KAIST
    $^2$ KAIST AI
}

\email{moonx010@kaist.ac.kr, kyudan@kaist.ac.kr, jchoo@kaist.ac.kr}


\usepackage{amsmath, amssymb, amsfonts, bm}
\usepackage{booktabs}
\usepackage{tikz}
\usetikzlibrary{positioning, arrows.meta, calc, fit, backgrounds}
\usepackage{multirow}
\usepackage{dblfloatfix}
\usepackage{kotex} 
\usepackage[table]{xcolor}
\begin{document}

\maketitle

\begin{abstract}
Real-world speech is often corrupted by multiple degradations simultaneously, including additive noise, reverberation, and nonlinear distortion.
Diffusion-based enhancement methods perform well on single degradations but struggle with compound corruptions.
Prior noise-aware approaches inject conditioning at the input layer only, which can degrade performance below that of an unconditioned model.
To address this, we propose injecting degradation conditioning, derived from a pretrained encoder with multi-task heads for noise type, reverberation, and distortion, into the timestep embedding so that it propagates through all residual blocks without architectural changes.
In controlled experiments where only the injection method varies, input-level conditioning performs worse than no encoder at all on compound degradations, while layer-wise injection achieves the best results.
The method also generalizes to diverse real-world recordings.
\end{abstract}

\section{Introduction}

As voice-based communication systems become pervasive, robust speech enhancement under real-world conditions remains a key challenge~\cite{richter2023speech, lu2022conditional, zhang2025urgent}.
In practice, speech signals are rarely degraded by a single source alone.
A phone call in a reverberant room, a voice recording captured on a low-quality device, or a conference call in a noisy environment all exhibit combinations of additive noise, reverberation, and nonlinear distortion~\cite{su2020hifi, liu2022voicefixer}.
These three categories can be viewed as representative of the principal signal corruption mechanisms: additive interference from environmental sources, convolutional effects from room acoustics~\cite{chiang2025dereverberation}, and nonlinear artifacts from recording devices or lossy transmission, whose prevalence is confirmed by recent universal enhancement challenges~\cite{saijo2025urgent, rong2025tsurgenet, sun2025scaling}.
While not exhaustive, this decomposition arguably covers the degradations most frequently encountered in practical speech communication settings.
Handling such compound degradations within a single model is therefore essential for practical deployment.

Diffusion-based generative models have emerged as a promising approach for speech enhancement, offering strong generalization through their iterative denoising process~\cite{richter2023speech, lu2022conditional, lay2025diffusion, han2025bridge}.
SGMSE+~\cite{richter2023speech} models clean speech distributions via score-based stochastic differential equations in the complex STFT domain, achieving state-of-the-art results on additive noise removal.
To further improve generalization, noise-aware methods inject external noise information as a conditioning signal to guide the enhancement process~\cite{hu2023nase, wang2023nadiffuse}.
NASE~\cite{hu2023nase} employs a pretrained BEATs~\cite{chen2023beats} audio encoder with a noise classification loss, adding the resulting embedding to the input spectrogram.
NADiffuSE~\cite{wang2023nadiffuse} similarly extracts noise representations and uses them as global conditions for the diffusion step.

However, these methods address only additive noise, leaving reverberation and nonlinear distortion unhandled despite their prevalence in real-world recordings~\cite{zhang2025urgent, su2020hifi, liu2022voicefixer}.
Furthermore, they rely on input-level conditioning injection, where the embedding is added at a single point in the network.
Since score networks such as NCSN++~\cite{song2020score} contain dozens of residual blocks, a single input-layer perturbation is progressively diluted through the network, leaving deeper layers unconditioned.

To address these limitations, we inject degradation conditioning into the timestep embedding of the NCSN++ backbone.
Since the timestep embedding is already consumed by every residual block, this naturally propagates the conditioning signal through the entire network without any architectural changes.
A pretrained WavLM~\cite{chen2022wavlm} encoder with three specialized heads jointly characterizes noise type, reverberation level, and distortion intensity via multi-task auxiliary losses~\cite{caruana1997multitask}, producing a single conditioning vector that unifies noise, reverberation, and distortion information to modulate all layers of the score network.
The multi-task design enables the encoder to disentangle co-occurring degradations, so that the score network receives informative conditioning even when multiple degradations are present simultaneously.
Through controlled experiments where only the injection method varies, we find that input-level conditioning performs worse than using no encoder at all on compound degradations, while our layer-wise injection achieves the best results across all metrics and generalizes to diverse real-world recordings.
Our contributions are twofold.
We reveal that shallow conditioning injection can hurt compound-degradation performance, and propose layer-wise conditioning via timestep embedding injection as a simple yet effective alternative for multi-degradation speech enhancement.

\section{Proposed Method}

\begin{figure*}[!t]
\centering
\resizebox{0.9\textwidth}{!}{%
\begin{tikzpicture}[
    >=Stealth,
    box/.style={draw, rounded corners=2pt, minimum height=0.85cm, align=center, font=\small},
    enc/.style={box, fill=encblue, minimum width=1.7cm},
    proj/.style={box, fill=projviolet, minimum width=1.6cm, minimum height=0.7cm, font=\footnotesize},
    head/.style={box, fill=headamber, minimum width=1.5cm, minimum height=0.7cm, font=\footnotesize},
    backbone/.style={draw, rounded corners=2pt, fill=bbteal,
        minimum width=2cm, minimum height=1.2cm, align=center, font=\small},
    emb/.style={box, fill=projviolet!40},
    op/.style={circle, draw, thick, inner sep=2pt, fill=white, font=\small},
    arr/.style={->, thick, draw=black!60},
    trunk/.style={thick, draw=black!60},
    dim/.style={font=\sffamily\scriptsize, text=black!50},
    loss/.style={font=\footnotesize},
]

\definecolor{encblue}{HTML}{B8D6FB}      
\definecolor{projviolet}{HTML}{C4B5FD}   
\definecolor{headamber}{HTML}{CDD2D9}    
\definecolor{bbteal}{HTML}{5FCEBD}       
\definecolor{bgacolor}{HTML}{E8F1FC}     
\definecolor{bgbcolor}{HTML}{DDF5EF}     
\definecolor{auxgray}{HTML}{8B9DB5}      
\definecolor{auxbg}{HTML}{E8EDF3}        


\node[inner sep=0pt] (audio) at (0, 0) {%
    \begin{tikzpicture}[scale=0.35]
        \draw[semithick, encblue!70!blue] plot[domain=0:2.4, samples=120]
            (\x, {0.5*sin(30*\x r)*exp(-0.8*(\x-1.2)*(\x-1.2)) + 0.15*sin(55*\x r)*exp(-1.2*(\x-0.8)*(\x-0.8)) + 0.08*rand});
        \node[font=\scriptsize, text=black!70] at (1.2, -1.05) {Noisy Audio};
    \end{tikzpicture}%
};
\node[enc] (wavlm) at (2.4, 0) {WavLM};
\node[enc, minimum width=1.6cm] (fp) at (4.7, 0) {Feature\\[-2pt]Projector};
\node[dim, anchor=north] at (fp.south) {$d_w \!\to\! d_h$};

\node[proj] (pn) at (7.4, 0.85) {Noise Proj};
\node[proj] (pr) at (7.4, 0.0)  {Reverb Proj};
\node[proj] (pd) at (7.4, -0.85) {Distort Proj};

\node[draw, dashed, thick, gray!40, rounded corners=5pt,
      inner xsep=8pt, inner ysep=5pt,
      fit=(pn)(pd),
      label={[font=\scriptsize\bfseries, text=black!60]above:Branch Projections}] (dbox) {};

\node[op, inner sep=1.5pt] (cat) at (9.2, 0) {\footnotesize$\Vert$};

\node[enc, minimum width=1.4cm] (mlp) at (10.5, 0) {MLP};
\node[dim, anchor=north] at (mlp.south) {$3d_b \!\to\! d$};

\node[emb] (temb) at (12.8, 1.2) {$\mathbf{t}_\text{emb}$};
\node[op] (plus) at (12.8, 0) {$+$};

\foreach \i in {2,1} {
    \fill[bbteal!40, draw=bbteal!60, rounded corners=2pt]
        ({13.8+\i*0.07}, {-0.6-\i*0.07})
        rectangle ({15.8+\i*0.07}, {0.6-\i*0.07});
}
\node[backbone] (res) at (14.8, 0) {ResBlocks};

\node[font=\small] (out) at (16.5, 0) {$\mathbf{s}_\theta$};
\node[loss, text=teal!70!black] (lscore) at (17.5, 0) {$\mathcal{L}_\text{score}$};

\node[box, fill=bbteal!30, font=\footnotesize] (stft) at (0, -3.8) {STFT};


\coordinate (aux_fan) at (6.0, -1.475);

\node[head] (nh) at (4.2, -2.1) {Noise Head};
\node[head] (rh) at (6.5, -2.1) {Reverb Head};
\node[head] (dh) at (8.8, -2.1) {Distort Head};

\node[loss, text=auxgray!80!black] (ln) at (4.2, -2.8)
    {$\mathcal{L}_\text{noise}$ (CE)};
\node[loss, text=auxgray!80!black] (lr) at (6.5, -2.8)
    {$\mathcal{L}_\text{reverb}$ (MSE)};
\node[loss, text=auxgray!80!black] (ld) at (8.8, -2.8)
    {$\mathcal{L}_\text{distort}$ (MSE)};

\node[font=\scriptsize\bfseries, text=black!60] at (10.6, -2.5)
    {Train-only};

\begin{scope}[on background layer]
    \draw[fill=bgacolor, draw=encblue!40, rounded corners=8pt]
        (1.2, 2.1) rectangle (11.3, -3.5);
    \draw[fill=bgbcolor, draw=teal!15, rounded corners=8pt]
        (11.8, 2.1) rectangle (16.3, -2.0);
    \node[fill=auxbg, draw=auxgray!60, rounded corners=6pt,
          dashed, dash pattern=on 4pt off 2pt,
          inner xsep=6pt, inner ysep=5pt,
          fit=(nh)(dh)(ln)(ld)] {};
\end{scope}

\node[font=\footnotesize\bfseries] at (6.25, 2.35) {(a) SLICE Encoder};
\node[font=\footnotesize\bfseries] at (14.05, 2.35) {(b) NCSN++ Score Network};

\node[op, inner sep=2pt, font=\scriptsize] (leg_add) at (15.6, -2.8) {$+$};
\node[font=\scriptsize, text=black!70, anchor=west] (leg_add_t) at (leg_add.east) {\;\,Addition};
\node[op, inner sep=2pt, font=\scriptsize] (leg_cat) at (15.6, -3.4) {$\Vert$};
\node[font=\scriptsize, text=black!70, anchor=west] (leg_cat_t) at (leg_cat.east) {\;\,Concatenation};
\coordinate (leg_br) at (leg_cat_t.south east);
\begin{scope}[on background layer]
    \node[fill=white, draw=gray!30, rounded corners=3pt,
          inner xsep=6pt, inner ysep=5pt,
          fit=(leg_add)(leg_cat)(leg_br)] {};
\end{scope}


\draw[arr] (audio.east) -- (wavlm.west);
\draw[arr] (audio.south) -- (stft.north);
\draw[arr] (wavlm.east) -- (fp.west);

\coordinate (fan) at (6.0, 0);
\draw[trunk] (fp.east) -- node[above, font=\scriptsize] {$\mathbf{h}$} (fan);
\draw[arr] (fan) |- (pn.west);
\draw[arr] (fan) -- (pr.west);
\draw[arr] (fan) |- (pd.west);

\draw[arr] (pn.east) -| (cat.north);
\draw[arr] (pr.east) -- (cat.west);
\draw[arr] (pd.east) -| (cat.south);

\draw[arr] (cat.east) -- (mlp.west);
\draw[arr] (mlp.east) -- (plus.west);
\draw[arr] (temb.south) -- (plus.north);
\draw[arr] (plus.east) -- (res.west);
\draw[arr] (res.east) -- (out.west);
\draw[arr, draw=black!30] (out.east) -- (lscore.west);

\draw[arr, rounded corners=5pt, draw=black!40] (stft.east) -| (res.south);


\draw[trunk, dashed, draw=black!60] (fan) -- (aux_fan);
\draw[arr, dashed, draw=black!60] (aux_fan) -| (nh.north);
\draw[arr, dashed, draw=black!60] (aux_fan) -| (rh.north);
\draw[arr, dashed, draw=black!60] (aux_fan) -| (dh.north);
\draw[arr, draw=auxgray!50] (nh.south) -- (ln.north);
\draw[arr, draw=auxgray!50] (rh.south) -- (lr.north);
\draw[arr, draw=auxgray!50] (dh.south) -- (ld.north);

\end{tikzpicture}%
}
\caption{Overview of SLICE.
\textbf{(a)}~The encoder produces~$\mathbf{h} \in \mathbb{R}^{d_h}$;
branch projections are concatenated and mapped to~$\mathbf{c}_\text{extra} \in \mathbb{R}^{d}$.
\textbf{(b)}~$\mathbf{c}_\text{extra}$ is added to~$\mathbf{t}_\text{emb}$
and propagated through every residual block.
Dashed orange: auxiliary heads (train-only).
$d_w\!=\!768$, $d_h\!=\!256$, $d_b\!=\!128$, $d\!=\!512$.}
\label{fig:architecture}
\end{figure*}
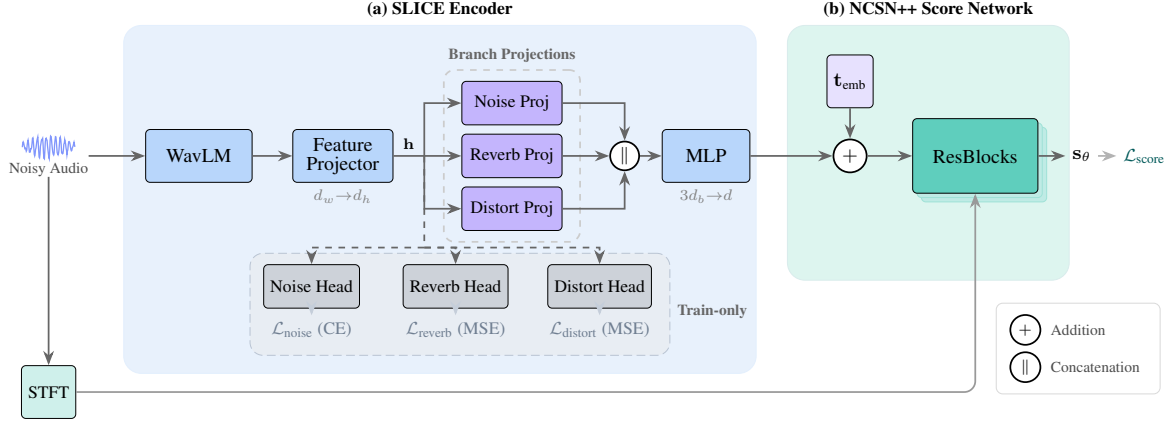

Our method extends the SGMSE+ framework with two components, as illustrated in Figure~\ref{fig:architecture}: (a)~a degradation-aware encoder that characterizes the corruption types present in the input signal, and (b)~a layer-wise conditioning mechanism that injects the resulting representation into every residual block of the score network via the timestep embedding.
We first review the SGMSE+ baseline, then describe each component in detail.

\subsection{Preliminaries: Score-Based Speech Enhancement}
\label{sec:prelim}

SGMSE+~\cite{richter2023speech} models the clean speech distribution conditioned on the noisy observation using a score-based SDE framework.
The forward process of the model gradually corrupts the clean signal $\mathbf{x}$ toward the noisy observation $\mathbf{y}$:
\begin{equation}
\mathrm{d}\mathbf{x}_t = \gamma(\mathbf{y} - \mathbf{x}_t)\,\mathrm{d}t + g(t)\,\mathrm{d}\mathbf{w}_t
\end{equation}
where $\gamma$ denotes the mean reversion rate and $g(t)$ is the diffusion coefficient.
A score network $\mathbf{s}_\theta(\mathbf{x}_t, \mathbf{y}, t) \approx \nabla_{\mathbf{x}_t} \log p_t(\mathbf{x}_t | \mathbf{y})$ is trained via denoising score matching, and clean speech is recovered by solving the reverse SDE with a predictor-corrector sampler.
The NCSN++~\cite{song2020score} backbone computes a timestep embedding $\mathbf{e}_t = \text{MLP}_\text{time}(t) \in \mathbb{R}^{d}$, where $d = 512$, that is added to features in every residual block, providing timestep awareness at all network depths.

\subsection{Multi-Degradation Encoder}
\label{sec:encoder}

We now describe the first component of our framework, the degradation-aware encoder, illustrated in Figure~\ref{fig:architecture}(a).
This module extracts a compact representation that captures the characteristics of each degradation type present in the input signal.

We employ WavLM-Base~\cite{chen2022wavlm}, a speech encoder pretrained on large-scale data to learn universal representations for speech processing tasks.
Given a degraded waveform $\mathbf{y}_\text{wav}$, WavLM produces frame-level features $\mathbf{F} = \text{WavLM}(\mathbf{y}_\text{wav}) \in \mathbb{R}^{T' \times d_w}$, where $d_w = 768$.
The encoder parameters are kept frozen throughout training to preserve the pretrained representations.
A convolutional post-processing network then compresses these semantically rich features into a shared representation $\mathbf{h} \in \mathbb{R}^{d_h}$, where $d_h = 256$, via mean pooling across the temporal dimension, retaining discriminative degradation information in a fixed-size descriptor suitable for downstream conditioning.

Prior noise-aware methods have observed that single-model performance exhibits trade-offs across different degradation types~\cite{hu2023nase, wang2023nadiffuse}, as a single representation that conflates all degradation types may fail to disentangle the distinct characteristics of each corruption.
Motivated by this observation, we design three specialized heads that independently characterize each degradation type, allowing the model to separately reason about the enhancement behavior for each corruption source.
Following the noise taxonomy of DEMAND~\cite{thiemann2013demand}, the noise head performs 11-class classification over 10 noise categories plus a ``none'' class.
The reverberation head regresses the room reverberation time $T_{60}$, while the distortion head estimates the nonlinear distortion intensity.
These heads serve dual purposes.
First, they provide auxiliary multi-task supervision~\cite{caruana1997multitask} that encourages the shared representation to learn discriminative features for each degradation type, mitigating the cross-degradation trade-offs that arise when a single encoder must handle all degradation types jointly.
Second, the per-head predictions enable interpretable degradation analysis at inference time, allowing the system to adapt its denoising behavior to the specific corruption profile of each input.

\subsection{Layer-wise Conditioning via Timestep Embedding Injection}
\label{sec:injection}

The second component of our framework, illustrated in Figure~\ref{fig:architecture}(b), injects the degradation representation into the score network.
We adopt existing timestep embedding mechanism of the NCSN++ backbone for this purpose.
The shared representation $\mathbf{h}$ is projected into three $d_b$-dimensional branch-specific embeddings, where $d_b = 128$, concatenated, and mapped to match the timestep embedding dimension:
\begin{align}
\mathbf{c}_k &= \mathbf{W}_k \mathbf{h}, \quad k \in \{\text{noise, reverb, distort}\} \label{eq:branch} \\
\mathbf{c}_\text{extra} &= \text{MLP}_\text{cond}([\mathbf{c}_\text{noise}; \mathbf{c}_\text{reverb}; \mathbf{c}_\text{distort}]) \label{eq:mlp}
\end{align}
where $[\cdot;\cdot]$ denotes concatenation and $\text{MLP}_\text{cond}: \mathbb{R}^{3d_b} \to \mathbb{R}^{d}$.
The conditioning is then injected by simple addition:
\begin{equation}
\tilde{\mathbf{e}}_t = \mathbf{e}_t + \mathbf{c}_\text{extra}
\label{eq:injection}
\end{equation}


Instead of relying on a single input-level injection such as NASE~\cite{hu2023nase}, this approach propagates the signal through all ${\sim}37$ residual blocks, making every layer of the score network degradation-aware. Crucially, this requires \textit{no architectural changes} to the backbone, as it only involves a single addition to the existing timestep embedding.

\definecolor{pastelblue}{HTML}{E7ECFE}

\begin{table*}[!t]
\centering
\caption{Speech enhancement results on noise-only (824 files) and multi-degradation (2472 files) conditions. All baselines are trained on noise-only data. Controlled ablations are in Table~\ref{tab:ablation}. Best per condition in \textbf{bold}. SDR denotes SI-SDR (dB).}
\label{tab:main_results}
\small
\setlength{\tabcolsep}{4pt}
\resizebox{0.62\textwidth}{!}{
\begin{tabular}{l|cccc|cccc}
\toprule
& \multicolumn{4}{c|}{\textbf{Noise-Only}} & \multicolumn{4}{c}{\textbf{Multi-Degradation}} \\
\textbf{Method} & PESQ & ESTOI & SDR & UTMOS & PESQ & ESTOI & SDR & UTMOS \\
\midrule
MP-SENet~\cite{lu2023mpsenet} & \textbf{3.16} & \textbf{0.88} & \textbf{19.6} & 3.90 & 2.39 & 0.62 & $-$0.3 & 3.00 \\
MetricGAN+~\cite{fu2021metricgan} & 3.13 & 0.83 & \phantom{0}8.6 & 3.63 & 2.53 & 0.62 & $-$5.5 & 2.82 \\
SGMSE+~\cite{richter2023speech} & 2.94 & 0.87 & 17.4 & 3.83 & 2.30 & 0.63 & \phantom{$-$}0.0 & 2.99 \\
SGMSE+ (dereverb)~\cite{richter2023speech} & 1.98 & 0.78 & 10.9 & 3.16 & 1.99 & 0.66 & $-$3.3 & 2.96 \\
NASE~\cite{hu2023nase} & 2.75 & 0.85 & 17.3 & 3.83 & 2.22 & 0.64 & $-$0.5 & 2.99 \\
\midrule
\rowcolor{pastelblue}\textbf{SLICE (Ours)} & 2.83 & 0.86 & 17.4 & \textbf{3.93} & \textbf{2.60} & \textbf{0.80} & \textbf{\phantom{$-$}3.7} & \textbf{3.71} \\
\bottomrule
\end{tabular}}
\end{table*}

\subsection{Training Objective}

The total loss combines the score matching objective with auxiliary multi-task losses:
\begin{equation}
\mathcal{L} = \mathcal{L}_\text{score} + \lambda \left( \mathcal{L}_\text{noise} + \mathcal{L}_\text{reverb} + \mathcal{L}_\text{distort} \right)
\label{eq:loss}
\end{equation}
where $\mathcal{L}_{\text{noise}}$ denotes the cross-entropy loss for noise classification, and $\mathcal{L}_{\text{reverb}}$ and $\mathcal{L}_{\text{distort}}$ represent the mean squared error (MSE) for regression. We set $\lambda = 0.3$ in all experiments.

Following classifier free guidance (CFG)~\cite{ho2022classifier}, each branch embedding is independently dropped to zero with probability $p = 0.1$ during training.
This enables the model to handle missing degradation types at inference, as it has learned to enhance speech with any subset of conditioning branches active.

\section{Experiments}

\subsection{Experimental Setup}

\subsubsection{Datasets}
We use VoiceBank-DEMAND~\cite{valentini2016investigating, thiemann2013demand} as the base dataset containing 11,572 training utterances with 10 noise types at SNRs of 0, 5, 10, and 15~dB.
To create multi-degradation training data, we augment each clean utterance with combinations of: (i) additive noise from DEMAND, (ii) reverberation via synthetic room impulse responses ($T_{60} \in [0.3, 1.0]$~s) generated with  \texttt{pyroomacoustics}~\cite{scheibler2018pyroomacoustics}, and (iii) nonlinear distortion via soft clipping $f(x) = \tanh(\alpha x)/\tanh(\alpha)$ with intensity $\alpha \in [1.5, 5.0]$.
The augmented dataset contains 34,716 utterances spanning six degradation categories.


We evaluate our method under three distinct conditions. The \textbf{Noise-only} test set comprises 824 utterances sourced from VoiceBank-DEMAND~\cite{voicebank_demand}. The \textbf{Multi-degradation} test set includes 2,472 files spanning six categories of degradation. For the \textbf{In-the-wild} evaluation, we utilize the VOiCES~\cite{richey2018voices} dataset featuring 1,600 room recordings, the DAPS~\cite{mysore_2014_4660670} dataset containing 1,500 real-device recordings, and the URGENT~\cite{saijo2025urgent} dataset with 1,000 diversely degraded files.

\subsubsection{Baselines}
We compare against five baselines trained on noise-only data: SGMSE+ (denoise)~\cite{richter2023speech}, SGMSE+ (dereverb)~\cite{richter2023speech}, MetricGAN+~\cite{fu2021metricgan}, MP-SENet~\cite{lu2023mpsenet}, and NASE~\cite{hu2023nase}.
To isolate the effect of the encoder and injection method, we train the SGMSE+ architecture from scratch on the same multi-degradation data without any encoder, as well as our full model but with NASE-style input addition instead of layer-wise injection.
These three models (no encoder, input addition, layer-wise injection) share identical training data, enabling a controlled comparison.

\subsubsection{Metrics and Implementation}
To align with recommendations for multifaceted evaluation beyond intrusive metrics~\cite{zhang2025urgent, deoliveira2025quality}, we report PESQ~\cite{rix2001perceptual}, ESTOI~\cite{jensen2016algorithm}, SI-SDR~\cite{le2019sdr}, and the neural MOS predictor UTMOS~\cite{saeki2022utmos}. Our model employs the NCSN++~\cite{song2020score} backbone coupled with the ordinary differential equation formulation from SGMSE+~\cite{richter2023speech}, while keeping the WavLM encoder frozen throughout the training process. The models are trained for 160 epochs utilizing the Adam optimizer with a learning rate of $10^{-4}$ and an exponential moving average decay of 0.999. We use a global batch size of 32, which is distributed across eight RTX 3090 GPUs with four samples allocated per device. Finally, the inference stage is conducted using 30 reverse steps.

\subsection{Main Results}

\textbf{Multi-degradation (fair comparison).}
Among models trained on the same data, the injection method is decisive.
Without any encoder, the SGMSE+ baseline already handles compound degradations reasonably, establishing a strong reference point.
Adding a WavLM encoder with \textit{input-level addition}, the injection used by NASE~\cite{hu2023nase}, \textit{significantly degrades} performance to ESTOI 0.73 and SI-SDR 1.4~dB, worse than using no encoder at all ($p < 10^{-7}$, paired $t$-test on 2,472 files) as shown in Table~\ref{tab:main_results}.
In contrast, the same encoder with \textit{layer-wise injection} significantly improves to ESTOI 0.80 ($p < 10^{-61}$) and SI-SDR 3.7~dB ($p < 10^{-15}$).
This demonstrates that layer-wise injection, not the encoder alone, is critical for leveraging degradation information in compound scenarios.

Baseline models trained on noise-only data achieve ESTOI between 0.62 and 0.66 with SI-SDR $\leq$0, performing substantially worse, and NASE trained on noise-only data achieves only ESTOI 0.64, further confirming that both multi-degradation data and layer-wise injection are necessary.

\textbf{Noise-only.}
On the standard noise-only benchmark, models designed exclusively for additive noise removal achieve higher PESQ~\cite{sun2025scaling, chao2025universal}, MP-SENet with parallel magnitude-phase denoising, and MetricGAN+~\cite{fu2021metricgan} which directly optimizes PESQ as its training objective.
Notably, our model produces the highest UTMOS across all models, surpassing even MP-SENet~\cite{lu2023mpsenet}, indicating superior perceptual quality despite the multi-degradation training objective.

\subsection{Ablation Studies}
\label{sec:ablation}

\begin{table}[!t]
\centering
\caption{Results of the ablation study evaluated on the multi-degradation test set. The term ``addition'' refers to the input-level addition method introduced by~\cite{hu2023nase}.}
\label{tab:ablation}
\small
\resizebox{0.9\columnwidth}{!}{%
\begin{tabular}{l|cccc}
\toprule
\textbf{Variant} & \textbf{PESQ} & \textbf{ESTOI} & \textbf{SDR} & \textbf{UTMOS} \\
\midrule
\textbf{SLICE} & \textbf{2.60} & \textbf{0.80} & \textbf{3.7} & \textbf{3.71} \\
\quad w/o aux.\ loss ($\lambda\!=\!0$) & 2.58 & 0.77 & 2.2 & 3.63 \\
\quad Input addition & 2.49 & 0.73 & 1.4 & 3.62 \\
No encoder & 2.60 & 0.77 & 2.3 & 3.70 \\
\quad Zero conditioning & 2.14 & 0.67 & $-$1.6 & 3.04 \\
\midrule
\quad Uniform weights & 2.60 & 0.80 & 3.7 & 3.71 \\
\bottomrule
\end{tabular}}
\end{table}

Table~\ref{tab:ablation} provides further ablations on the multi-degradation test set.
All variants share the same training data and WavLM encoder, except ``No encoder'', consistent with the controlled comparison in Table~\ref{tab:main_results}.

\textbf{Layer-wise vs.\ shallow injection.}
As also shown in Table~\ref{tab:main_results}, replacing our layer-wise injection with input addition~\cite{hu2023nase}, where the conditioning vector is added only once to the input STFT representation, degrades every metric and performs even worse than using no encoder at all.
We hypothesize two mechanisms: (1) at the input layer, the added embedding shifts the complex STFT representation in a way that disrupts learned spectrogram processing of the network, and (2) this single perturbation is progressively diluted through ${\sim}37$ residual blocks, so deeper layers receive no conditioning signal.
Layer-wise injection avoids both issues by modulating affine transform of every layer via the timestep embedding, without altering the input spectrogram.
This is consistent with recent findings that different layers in deep speech enhancement networks play distinct roles and benefit from dedicated conditioning~\cite{parvathala2025dynamic}.

\textbf{Multi-task loss.}
Setting the auxiliary loss weight $\lambda$ to 0 reduces ESTOI from 0.80 to 0.77 and UTMOS from 3.71 to 3.63.
Per-degradation analysis reveals this effect concentrates on reverberant conditions, for noise+reverb+distortion, ESTOI drops from 0.647 to 0.517 and SI-SDR from $-$18.4 to $-$23.8~dB without auxiliary losses.
This confirms that multi-task supervision strengthens reverb discrimination of the encode.

\textbf{Zero conditioning.}
Setting $\mathbf{c}_\text{extra}$ to $\mathbf{0}$ at inference causes large degradation, dropping PESQ from 2.60 to 2.14 and UTMOS from 3.71 to 3.04, confirming that the model relies heavily on the conditioning signal.

\textbf{Adaptive vs.\ uniform weighting.} Interestingly, dynamically scaling each branch using the predicted confidence of encoder yields no performance gain over simply applying uniform weights. This outcome strongly suggests that our layer-wise injection mechanism naturally equips the score network to figure out which branches matter most. As a result, adding explicit weighting during inference is an unnecessary step, as the network already manages this internally.

\subsection{Analysis}

\textbf{Encoder head accuracy.}
The multi-task encoder demonstrates an ability to characterize various audio degradations. Specifically, it achieves a noise binary detection accuracy of 96.7\%, while maintaining a high $T_{60}$ correlation of 0.981 with a minimal mean absolute error of 0.033 for reverberation. Furthermore, it reaches a distortion intensity correlation of 0.845 alongside a solid detection accuracy of 86.2\%. These strong metrics highlight how the auxiliary losses shape a well-calibrated encoder that can accurately pinpoint each type of degradation.
\begin{table}[!t]
\centering
\caption{Detailed performance breakdown across six distinct degradation categories within the multi-degradation test set.}
\label{tab:perdeg}
\small 
\setlength{\tabcolsep}{3.5pt}
\resizebox{0.45\textwidth}{!}{
\begin{tabular}{l|c|cccc}
\toprule
\textbf{Type} & \textbf{N} & \textbf{PESQ} & \textbf{ESTOI} & \textbf{SDR} & \textbf{UTMOS} \\
\midrule
Noise only           & 824 & 2.82 & 0.86 & 17.5  & 3.93 \\
Noise+Reverb         & 330 & 1.72 & 0.65 & $-$17.2 & 3.31 \\
Noise+Distort        & 318 & 2.84 & 0.86 & 15.8  & 3.94 \\
Noise+Reverb+Distort & 338 & 1.74 & 0.65 & $-$18.4 & 3.34 \\
Reverb only          & 341 & 2.01 & 0.74 & $-$17.1 & 3.42 \\
Distort only         & 321 & 4.21 & 0.98 & 23.3  & 4.05 \\
\bottomrule
\end{tabular}}
\end{table}

\textbf{Per-degradation breakdown.}
As detailed in Table~\ref{tab:perdeg}, the model exhibits varying performance across different degradation types. It handles distortion nearly perfectly, achieving a PESQ of 4.21 and a UTMOS of 4.05. Conversely, reverberation leads to a noticeable degradation in SI-SDR. However, the UTMOS remains above 3.3 even under these challenging reverberant conditions, indicating that the perceptual quality is still reasonably preserved despite the low SI-SDR. This represents a meaningful enhancement, especially considering that single-degradation baselines fare far worse in reverberant environments with ESTOI scores hovering between 0.20 and 0.30.
\begin{table}[!t]
\centering
\caption{Evaluation on in-the-wild dataset. SGMSE+ (pretrained) is the publicly available denoise model~\cite{richter2023speech}; SGMSE+ (no enc.) is trained from scratch on the same multi-degradation data as SLICE but without an encoder. SI-SDR is omitted as it is unreliable for reverberant recordings.}
\label{tab:wild}
\small
\resizebox{0.45\textwidth}{!}{
\begin{tabular}{ll|ccc}
\toprule
Dataset & Model & PESQ & ESTOI & UTMOS \\
\midrule
\multirow{3}{*}{VOiCES} & SGMSE+ (pretrained) & 1.20 & 0.34 & 1.98 \\
& SGMSE+ (no enc.) & \textbf{1.59} & 0.65 & \textbf{2.83} \\
& \textbf{SLICE} & 1.56 & \textbf{0.67} & 2.77 \\
\midrule
\multirow{3}{*}{DAPS} & SGMSE+ (pretrained) & 2.10 & \textbf{0.69} & 2.96 \\
& SGMSE+ (no enc.) & 2.04 & 0.65 & 2.96 \\
& \textbf{SLICE} & \textbf{2.20} & 0.60 & \textbf{3.32} \\
\midrule
\multirow{3}{*}{URGENT} & SGMSE+ (pretrained) & \textbf{1.68} & \textbf{0.70} & 2.97 \\
& SGMSE+ (no enc.) & 1.66 & 0.67 & 3.01 \\
& \textbf{SLICE} & 1.67 & 0.68 & \textbf{3.10} \\
\bottomrule
\end{tabular}}
\end{table}

\textbf{In-the-wild generalization.}
Table~\ref{tab:wild} evaluates the generalization capabilities of three models on real-world, unseen recordings: SLICE, SGMSE+ (no enc.) trained from scratch on multi-degradation data, and the publicly available SGMSE+ pretrained on noise-only data. 
Both multi-degradation models dramatically outperform the pretrained SGMSE+ on the VOiCES dataset, confirming that diverse training data is the primary driver of wild-data generalization. 
Furthermore, SLICE and SGMSE+ (no enc.) achieve comparable PESQ and ESTOI scores across all three datasets, suggesting that the benefit of the encoder is less pronounced on out-of-domain, real-world recordings compared to the controlled test set. 
Regarding perceptual quality, the UTMOS results present a more nuanced story: the proposed model achieves the highest scores on DAPS and URGENT, while SGMSE+ (no enc.) performs slightly better on VOiCES. 
Overall, both multi-degradation models consistently and substantially outperform the noise-only pretrained baseline in terms of perceptual quality.

\section{Conclusion}
Through controlled experiments isolating the injection method, we demonstrated that the depth of conditioning injection plays a significant role in the effectiveness of degradation-aware conditioning for score-based speech enhancement. Shallow input addition, a common approach in prior work, can perform less effectively than unconditioned models on compound degradations. In contrast, injecting the same conditioning into the timestep embedding yields more robust results. Furthermore, a WavLM encoder utilizing multi-task auxiliary losses produces well-calibrated degradation representations, enabling a single model to jointly handle noise, reverberation, and distortion. These results indicate that the mere presence of conditioning does not guarantee performance improvements. We suggest that the method of injection is as important as the conditioning feature itself, presenting a finding with potential implications for conditional score-based models beyond speech enhancement.

\section{Generative AI Use Disclosure}
Generative AI tools were used for code development assistance and manuscript editing.
All experimental design, analysis, and scientific contributions are the work of the authors.

\bibliographystyle{IEEEtran}
\bibliography{mybib}

\end{document}